\documentclass[cmfonts,a4paper]{aipproc}
\layoutstyle{8d}
\usepackage{amssymb,amsmath,amsthm,amscd,verbatim}
\usepackage[mathcal]{euler}
 \font\palba=pplb at 20pt
 \font\palo=pplr at 14pt
 \font\paloa=pplr at 14pt
 \font\palom=pplr at 12 pt
 \font\palomar=pplr at 10 pt

\title{\palba Maximum Entropy method with non-linear moment constraints: challenges}

\author{\palo M. Grendar, Jr. and \underline{M. Grendar}}
{
 address = {Institute of Mathematics and Computer Science of Mathematical Institute of
 Slovak Academy of Sciences (SAS) and of Matej Bel University, Severn\'a ulica 5, 974 00
        Bansk\'a Bystrica, Slovakia \&
 Institute of Measurement Science of SAS,
           D\'ubravsk\'a cesta 9, 841 04 Bratislava,  Slovakia. umergren@savba.sk
           }}

\def\vc{\boldsymbol}
\def\vcs{\boldsymbol}

\begin{document}
\begin{abstract}
Traditionally, the Method of (Shannon-Kullback's) Relative Entropy
Maximization (REM) is considered with linear moment constraints. In this
work, the method is studied under frequency moment constraints which are
non-linear in probabilities. The constraints challenge some justifications
of REM since a) axiomatic systems are developed for classical linear moment
constraints, b) the feasible set of distributions which is defined by
frequency moment constraints  admits several entropy maximizing
distributions ($I$-projections), hence probabilistic justification of REM
via Conditioned Weak Law of Large Numbers cannot be invoked. However, REM
is not left completely unjustified in this setting, since Entropy
Concentration Theorem  and Maximum Probability Theorem can be applied.

Maximum R\'enyi/Tsallis' entropy method (maxTent) enters this work because
of non-linearity of $X$-frequency moment constraints which are used in
Non-extensive Thermodynamics. It is shown here that under $X$-frequency
moment constraints maxTent distribution can be unique and different than the
$I$-projection. This implies that maxTent does not choose the most probable
distribution and  that the maxTent distribution  is asymptotically
conditionally improbable. What are adherents of maxTent accomplishing when
they maximize R\'enyi's or Tsallis' entropy?
\end{abstract}

\maketitle

\section{\paloa 1 Introduction}

Let $\mathcal{\Pi}$ be a set of empirical probability mass functions (types)
which are defined on $m$-element support and which can be based on random
samples of size $n$. Let the supposed source  of the types be probability
mass function $\vc q$. A problem (from category of ill-posed inverse
problems) of recovering probability distribution from $\mathcal{\Pi}$
amounts to  selection of type(s) from $\mathcal{\Pi}$, in particular when $n
\rightarrow \infty$.

The problem (called hereafter Boltzmann-Jaynes Inverse Problem, BJIP) can be
met in many branches of science, ranging from Statistical Physics (where it
originated) to Computer Tomography. Several approaches to the problem can
be found in the literature. While most of them are tailored to needs of the
particular branch of science, the method of (Shannon-Kullback's) Relative
Entropy Maximization (REM) is considered as the general solution to the
problem by mathematicians. Arguments which justify application of REM for
selection of distribution from $\mathcal\Pi$ in BJIP range from axiomatic,
through probabilistic and game-theoretic to pragmatic, and others. As rule,
in order to be valid they put certain requirements 
on  $\mathcal\Pi$ and $n$.

So far, most of the REM-justifying work concentrated on the case of
$\mathcal\Pi$  defined by the usual linear moment constraints. Such
$\mathcal\Pi$ possesses the attractive property of convexity, which thanks
to concavity of the Shannon-Kullback's entropy implies uniqueness of
REM-selected distribution (called $I$-projection of $\vc q$ on
$\mathcal\Pi$, in the Information Theory). Linearity of the constraints lays
behind the well-known exponentiality of the $I$-projection.

As  \cite{Romera}  indicates, so-called frequency moment constraints appear
rather naturally in several places in Physics. Frequency moments are
non-linear in probabilities and the feasible set  $\mathcal\Pi_f$ which
they define is non-convex. Due to the non-linearity and a symmetry of the
constraints, there are multiple $I$-projections of $\vc q$ on
$\mathcal\Pi_f$. The non-linearity of  moments, non-convexity of the
feasible set, non-exponentiality of recovered distribution and its
non-uniqueness challenge several justifications of REM. Two of the most
widely employed REM-justifying arguments: axiomatizations and Conditioned
Weak Law of Large Numbers  cannot be invoked in this setting since
axiomatic systems are developed for linear constraints and CWLLN requires
assumption of uniqueness of $I$-projection. Is there then  any reason to
select the most entropic distribution from $\mathcal\Pi_f$? Yes, since
Entropy Concentration Theorem (ECT) and Maximum Probability  Theorem (MPT)
can be readily used to justify MaxEnt also in this case. Though MPT was
originally stated  with unique $I$-projection in mind, the Theorem can be
instantly extended also to the case of multiple $I$-projections.

The frequency moment constraints can be viewed as a special case of Tsallis'
(cf. \cite{Tsallis}) or MNNP (cf. \cite{TMP}, \cite{MNPP}) constraints
which are used in 'hot topic' Non-extensive Thermodynamics (NET). The
constraints are as well non-linear in probabilities. NET has arisen from
Tsallis' prescription to select from set which the constraints define such a
distribution which maximizes Tsallis' entropy.  Thus, in this area REM was
displaced (or generalized, if you wish) by maximization of Tsallis'
entropy. Besides axiomatic justifications (which are based on extensions of
those of REM) and declared success of maxTent in modeling power-law
phenomena (which allegedly REM cannot model), there is however yet no
probabilistic justification of the method.

The paper is organized as follows: First, the necessary terminology and
notation is set down. Then probabilistic justifications of REM: CWLLN, ECT
and MPT are reviewed from perspective of their applicability in the case of
multiple $I$-projections. Maximum Probability Theorem is stated in the
general form which covers the situation of multiple $I$-projections. Also,
applicability of other justifications is briefly discussed. Next we turn to
the simplest of non-linear moment constraints: frequency moment constraints
and note that $I$-projection on $\mathcal\Pi_f$ is non-unique and
non-exponential. Frequency moments constraints are then used to provide an
illustration for the general form of Maximum Probability Theorem. Next,
Tsallis' and R\'enyi's entropies are introduced, and it is noted that under
frequency moment constraints maximization of R\'enyi-Tsallis' entropy
(maxTent) selects no distribution. Under MNNP constraints it does, but as
it will be shown, the maxTent-selected distribution can be unique but
different than the $I$-projection. Consequences of this finding for maxTent
are discussed. Concluding comments sum up the paper and point to further
considerations. Appendix describes a method for finding $I$-projections on
$\mathcal\Pi_f$.

\section{\paloa 2 Terminology and notation}

Let $\mathcal{X} \triangleq \{x_1, x_2, \dots, x_m\}$ be a discrete finite
set called support, with $m$ elements and let $\{X_l, l = 1, 2, \dots, n\}$
be a sequence of size $n$ of identically and independently drawn random
variables taking values in $\mathcal{X}$.

A type $\vcs\nu \triangleq [n_1, n_2, \dots, n_m]/n$ is an empirical
probability mass function which can be based on sequence $\{X_l, l = 1, 2,
\dots, n\}$. Thus, $n_i$ denotes number of occurrences of $i$-th element of
$\mathcal{X}$ in the sequence.

Let $\mathcal{P(X)}$ be a set of all probability mass functions (pmf's) on
$\mathcal{X}$. Let $\mathcal\Pi \subseteq \mathcal{P(X)}$.

Let the supposed source of the sequences (and hence also of types) be $\vc
q \in \mathcal{P(X)}$, called (prior) generator.

Let ${\pi}(\vcs\nu)$ denote the probability that  $\vc q$ will generate
type $\vcs\nu$, ie. ${\pi}(\vcs\nu) = \frac{n!}{n_1!\,n_2!\,\dots\, n_m!}
\prod_{i =1}^m q_i^{n_i}$. Then, ${\pi}(\vcs\nu \in \mathcal{A})$ denotes
the probability that $\vc q$ will generate a type $\vcs\nu$ which belongs to
$\mathcal{A} \subseteq \mathcal\Pi$, ie. ${\pi}(\vcs\nu \in \mathcal{A}) =
\sum_{\vcs\nu \in \mathcal{A}} {\pi}(\vcs\nu)$. Finally, let ${\pi}(\vcs\nu
\in \mathcal{A} | \vcs\nu \in \mathcal{\Pi})$ denote the conditional
probability that if $\vc q$ generates type $\vcs\nu \in \mathcal\Pi$ then
the type belongs to $\mathcal{A}$. It is assumed that the  conditional
probability exists.

$I$-projection  $\hat{\vc p}$ of $\vc q$ on  set $\mathcal{\Pi} \subseteq
\mathcal{P(X)}$ is such $\hat{\vc p} \in \mathcal{\Pi}$ that $I(\hat{\vc
p}\| \vc q) = \inf_{\vc p \in \mathcal{\Pi}} I(\vc p\|\vc q)$,
where\footnote{There, $\log 0 = - \infty$, $\log \frac{b}{0} = + \infty$,
$0 \cdot (\pm\infty) = 0$, conventions are assumed.  Throughout the paper
$\log$ denotes the natural logarithm.} $I(\vc p\| \vc q) \triangleq
\sum_{\mathcal X} p_i \log \frac{p_i}{q_i}$ is the $I$-divergence.
$I$-divergence is  known under various other names: Kullback-Leibler's
distance, KL number, Kullback's directed divergence, etc. When taken with
minus sign it is known as (Shannon-Kullback's) relative entropy.

General framework of this work is established by Boltzmann-Jaynes inverse
problem (BJIP)\footnote{Equivalently the framework could be phrased as a
problem of induction (or updating), cf. \cite{ggbayes}.}:

{\it Let there be a set $\mathcal\Pi \subseteq \mathcal{P}(\mathcal{X})$ of
types which are defined on $m$-element support $\mathcal{X}$ and which can
be based on random samples of size $n$. Let the supposed source of the
random samples (and thus also types) be  $\vc q$. BJIP amounts to selection
of specific type(s) from $\mathcal\Pi$ when information $\{\mathcal{X}, n,
\vc q, \mathcal\Pi\}$ is supplied.}

{\bf Example 1}: Let $n=6$, $\mathcal{X} = [1\ \;2\ \;3]$, $\vc q = [1/3\
\;1/3\ \;1/3]$ and let the feasible set comprise all such types which have
probability of one of the support-points equal to 2/3, ie. $\mathcal\Pi
=\{[2/3\ \,1/6\ \,1/6], [2/3\  \,1/3\ \,0], \dots\}$ where the dots stand
for the remaining 7 permutations of the two listed types. Given the
information $\{n, \mathcal{X}, \vc q, \mathcal\Pi\}$ the BJIP task is to
select a type from the set $\mathcal\Pi$. $\qquad\qed$

If $\mathcal{\Pi}$ contains more than one type (as it is the case in the
above Example), the BJIP becomes under-determined and in this sense
ill-posed.

\section{\paloa 3 Justifications of REM}

\subsection{\palom 3.1 Conditioned Weak Law of Large Numbers}

A result of the Method of Types, which was developed in the Information
Theory (cf. \cite{CsiszarMT}), provides a probabilistic justification for
application of REM method  for solving BJIP, when $n$ tends to infinity and
$\mathcal{\Pi}$ has certain properties. The result is usually known as
Conditioned Weak Law of Large Numbers (CWLLN), or as Gibbs conditioning
principle (in large deviations literature, see \cite{Ellis}, \cite{DZ}).
The argument shows (loosely speaking) that any type from $\mathcal\Pi$
which is generated by $\vc q$ and is not close (in $L_1$-norm) to the
$I$-projection of $\vc q$ on $\mathcal\Pi$ becomes conditionally improbable
to come from $\vc q$ as sample size grows large. To establish this result
(cf. \cite{Vasicek}, \cite{VC}, \cite{G}, \cite{sCsiszar}, \cite{CT},
\cite{LS2}, \cite{LN}, \cite{LPS}) assumption of uniqueness of
$I$-projection is needed.

{\bf (CWLLN)} {\it Let  $\hat{\vc p}$ be unique $I$-projection of $\vc q$ on
$\mathcal\Pi$. Let  $\vc q \notin \mathcal{\Pi}$. Then for any $\epsilon
> 0$}

\begin{align}
\lim_{n\rightarrow\infty} \pi\left(\left|\nu_i - \hat p_i\right| > \epsilon
\,\,\left\vert\,\, \vcs\nu \in \mathcal{\Pi}\right.\right) &= 0 & i &= 1, 2,
\dots, m
\end{align}

Well-studied is the case of closed, convex $\Pi$, which ensures uniqueness
of $I$-projection, provided that it exists (cf. \cite{sCsiszar}, and
\cite{LS2}, \cite{LN}, \cite{LPS} for further developments). As it is
well-known, in this case the $I$-projection belongs to the exponential
family of distributions (see \cite{gCsiszar}).

\subsection{\palom 3.2 Entropy Concentration Theorem}

Without the assumption of  uniqueness of the $I$-projection, a  claim known
as the Entropy Concentration Theorem (ECT), weaker than (1), can be still
made (see \cite{CT}):

\smallskip

{\bf (ECT)} {\it Let  $\mathcal{\Pi} \subseteq \mathcal{P(X)}$ be nonempty.
Let $\hat{I}$ be such that $\hat{I} \le I(\vcs\nu \| \vc q)$ for any
$\vcs\nu \in \mathcal{\Pi}$. Then for any $\epsilon > 0$}
\begin{equation}
\lim_{n \rightarrow \infty}  \pi\left(\left| I\left(\vcs\nu \| \vc q\right)
- \hat{I}\right| < \epsilon\,  \left |\right. \vcs\nu \in
\mathcal{\Pi}\right) = 1
\end{equation}

Assumption (of whatever form) which guarantees existence and uniqueness of
the $I$-projection is crucial for coming from statement (2) to the stronger
claim (1).

\subsection{\palom 3.3 Maximum Probability  Theorem}

Maximum Probability Theorem (MPT), which was originally (see \cite{ggwhat},
Thm 1.) stated with unique $I$-projection in mind, claims that the type
$\hat{\vcs\nu}$ in $\mathcal\Pi$ which the (prior) generator $\mathbf q$ can
generate with the highest probability converges to the $I$-projection of
$\vc q$ on $\mathcal \Pi$, as $n \rightarrow\infty$. However proof of the
Theorem (cf. \cite{ggwhat}) covers more general situation of multiple
$I$-projections and thus allows to state MPT in the following general form:

{\bf (MPT)} {\it  Let ${\mathbf{q}}$ be a generator. Let {\rm
differentiable\/} constraint $F({\vcs{\nu}}) = 0$ define feasible set of
types $\mathcal\Pi_n \subseteq \mathcal\Pi$ and let $\mathcal\Pi \triangleq \{\vc p: F({\vc{p}})
= 0\}$ be the corresponding feasible set of probability mass functions. Let
$\hat{\vcs\nu}_j \triangleq \arg\, \max_{\vcs\nu \in \mathcal\Pi_n}
\pi(\vcs\nu)$, $j = 1, 2, \dots, l$, be types which have among all types
from $\mathcal\Pi_n$ the highest probability of coming  from the
generator $\vc q$. Let there be $k$ $I$-projections $\hat{\vc p}_1,
\hat{\vc p}_2, \dots, \hat{\vc p}_k$ of $\vc q$ on $\mathcal\Pi$. And let
$n\rightarrow\infty$. Then $l=k$ and $\hat{\vcs\nu}_j = \hat{\mathbf{p}}_j$
for $j=1, 2, \dots, k$. }

It should be noted that MPT argument implies that REM is only a special,
asymptotic form of simple and self-evident method (called  Maximum
Probability method (MaxProb) at \cite{ggwhat}) which seeks in $\mathcal\Pi$
such types which the  generator $\vc q$ can generate with the highest
probability. Thus applicability of REM in BJIP is inherently limited to the
case of sufficiently large $n$.

Also, it is worth noting that a bayesian interpretation can be given to
MaxProb, which thanks to MPT carries over into REM/MaxEnt (cf.
\cite{ggbayes}).

From the perspective of the current work, it is important that the MPT
holds also when the feasible set  admits multiple types with
the highest value of the probability $\hat{\pi}$. An illustration of the
convergence of most probable types to $I$-projections will be given  in the
Section 4, where such a set is determined by  frequency
moment constraints.

\subsection{\palom 3.4 Axiomatic systems}

Besides the probabilistic arguments several axiomatic a\-p\-pro\-ach\-es
were developed to support  maximization of Shannon's entropy or relative
entropy as the only logically consistent method for solving
BJIP\footnote{Strictly speaking, the axiomatizations assume BJIP with
either $n$ unknown or $n$ bigger than any limit. They seem to be
inappropriate for BJIP with finite sample size.}. However, it should be
noted that maximization of  R\'enyi's entropy was as well found to satisfy
some of the axiomatic systems, which had been developed to justify REM (see
\cite{Uffink}). For purposes of the presented work it is sufficient to note
that the axiomatic system (cf. \cite{aCsiszar}) which is perhaps the most
widely accepted requires assumption of linearity of the constraints (or, in
general, convexity of $\mathcal\Pi$). A non-axiomatic argument based on
potential-probability density relationship and a complementarity (cf.
\cite{ggwhy}) is restricted to the linear constraints as well. Also a
game-theoretic view of REM (see \cite{Topsoe}) assumes  the linear
constraints.

To sum up: When $\mathcal\Pi$ admits several $I$-projections the
justifications of REM which are readily available reduce to Entropy
Concentration Theorem and  Maximum Probability Theorem.

\section{\paloa 4 Frequency moment constraints}

This study was triggered by an interesting paper by Romera, Angulo and
Dehesa (cf. \cite{Romera}) on frequency moment problem. There also links to
statistical considerations of the frequency moments as well as to their
applications in Physics can be found.

In the simplest case of single frequency moment constraint, feasible set of
types is defined as $\mathcal{\Pi}_f \triangleq \{\vc p: \sum_{i=1}^m
p_i^\alpha - a = 0, \sum_{i=1}^m p_i - 1 = 0\}$, where $\alpha, a \in
\mathbf{R}$. If $m > 2$, the problem of selection of type becomes
ill-posed. Note that the first constraint is for $\alpha \neq 1$ non-linear
in $\vc p$ and $\mathcal{\Pi}_f$ is non-convex.

\subsection{\palom 4.1 $I$-projection: non-uniqueness and non-exponentiality}

It is straightforward to observe that $I$-projection of $\vc q$ on
$\mathcal\Pi_f$ possesses  a symmetry, in the sense that if certain
$\hat{\vc p}$ is $I$-projection of $\vc q$ on ${\mathcal\Pi}_f$ then any
permutation of the vector  $\hat{\vc p}$  should necessarily be also
$I$-projection.

Within this Section $\vc q$ will be assumed uniform (for a reason which is
implied by discussion at Section 5.1), denoted $\vc u$. Note that when
uniform generator is assumed, the method of Relative Entropy Maximization
reduces to Maximum Shannon's Entropy method (abbreviated usually MaxEnt).

The non-convexity of feasible set makes the problem of maximization of
Shannon's entropy analytically unsolvable.  
Critical value of $p_i$ is expressed
as: $p_i(\lambda) = k(\lambda) e^{-\lambda \alpha p_i^{\alpha - 1}}$, where
$k(\lambda) = \sum e^{-\lambda \alpha p_i^{\alpha - 1}}$. Note that the
expression is explicitly self-referential.

Thus, the $I$-projections should be searched out either numerically or by a
method which is described at the Appendix.

\subsection{\palom 4.2 MaxProb justification of REM: multiple $I$-projections}

That the most probable types indeed converge to the corresponding
$I$-projections as the general form Maximum Probability Theorem states
 will be illustrated by the following Example.

{\bf Example 2}:\ \  Let $\alpha = 2$, $\mathcal{X}= [1\ \,2\ \,3]$, $m =
3$ and $a = 0.42$ (the value was obtained for $p = [0.5\ 0.4\ 0.1]$).

For $n = 10, 30, 330, 1000, 2000$ the feasible sets $\mathcal{\Pi}_f$ were
constructed. For example, $\mathcal{\Pi}_{f,10}$ contains $\vcs\nu = [5\ 4\
1]/10$ and all its permutations (ie. $[5\ 1\ 4]/10$, etc). This will be
called {\it group of types}. $\mathcal{\Pi}_{f,30}$ contains two groups:
$[15\ \,12\ \,3]/30$ and $[17\ 8\ 5]/30$. The last one has higher
probability of coming from uniform prior generator. For $n=330$ the
feasible set comprises groups $[0.0939\ 0.4333\ 0.4727]$, $[0.5666\ 0.2666\
0.1666]$, $[0.1\ 0.4\ 0.5]$ and the group $[0.1939$ $0.2333\ 0.5727]$,
which has the highest probability of being generated by $\vc u$.

For each $n$, among the feasible types, the most probable $\hat{\vcs\nu}$
which could be drawn from the uniform prior generator was picked up. They
are stated at the Table 1 together with a corresponding $I$-projection of
$\vc u$ on $\mathcal\Pi_f$.

\begin{table}[h!]
\caption{The most probable type, for growing $n$.}
\begin{tabular}{ c  c  c  c }
\hline
   $n$ &     &  $\hat{\vcs\nu}|\vc u$\\
\hline
   10 &        0.1  &  0.4  &  0.5 \\
   30 &        0.166 & 0.266 & 0.566 \\
  330 &        0.1939 & 0.2333 & 0.5727 \\
  1000 &        0.1990 & 0.2280 & 0.5730 \\
  2000 &       0.2080 & 0.2185 & 0.5735 \\
\hline
 $\hat{\vc p}$ &          0.2131 & 0.2131 & 0.5737 \\
\hline
\end{tabular}
\end{table}

Clearly, the most probable type (hence also the whole permutation group of
6 most probable  types) converges to the pmf (permutation group of 3 pmf's)
which maximizes Shannon's entropy. \qed

\subsection{\palom 4.3 maxTent: no selection}

At this point, both R\'enyi's and Tsallis' entropies will be introduced.
R\'enyi's entropy (cf. \cite{S}, \cite{Renyi}) is defined as $H_{R}(\vc p)
\triangleq \frac{1}{1-\alpha} \log\left(\sum_{i=1}^m p_i^\alpha\right)$,
where $\alpha \in \mathbf{R}$, $\alpha \neq 1$.

Tsallis' entropy $H_T$ (cf. \cite{HC}, \cite{Vajda}, \cite{Tsallis})
 is linear approximation of R\'enyi's entropy:
$H_{T}(\vc p) \triangleq \frac{1 - \sum_{i=1}^m p_i^\alpha}{\alpha-1}$,
where $\alpha \in \mathbf{R}$, $\alpha \neq 1$.

R\'enyi's entropy attains its maximum at the same pmf as does Tsallis'
entropy. Thus, hereafter maxTent will denote both method of maximum
R\'enyi's and Tsallis' entropy at once.  maxTent will be discussed in
greater detail in Section 5. Here it suffices to note that in the set
$\mathcal{\Pi}_f$ which is defined by the frequency moment constraint each
type has the same value of R\'enyi's (or Tsallis') entropy. In other words,
maxTent refuses to make a choice from ${\mathcal\Pi}_f$. Recall that MaxEnt
selects  $I$-projections, and ECT implies that types conditionally
concentrate on the $I$-projections in such a way, that as $n$ gets large
there is virtually no chance to find a type which has value of Shannon's
entropy different than the maximal one. MPT complements it by stating that
most probable types turn into the $I$-projections, as $n$ goes to infinity.

\section{\paloa 5 $X$-frequency moment constraints}

Frequency moment constraints can be viewed as a special case of non-linear
constraints which were originally introduced into Statistical Mechanics by
Tsallis (see \cite{Tsallis}). Tsallis' constraints define feasible set
$\mathcal{\Pi}_T$ as follows: $\mathcal{\Pi}_{T} \triangleq \{\vc p:
\sum_{i=1}^m p_i^\alpha x_i - a = 0, \sum_{i=1}^m p_i - 1 = 0\}$.

Tsallis' constraints were for Physics reasons superseded by TMP constraints
(see \cite{TMP}). Later on, the TMP constraints were rearranged by Mart\'\i
nez, Nicol\'as, Pennini and Plastino \cite{MNPP} in
 MNPP form which allows for simpler analytic
tractability. The TMP constraints in MNNP form specify feasible set  as
$\mathcal{\Pi}_{\tau} \triangleq \{\vc p: \sum_{i=1}^m p_i^\alpha(x_i  - b)
= 0, \sum_{i=1}^m p_i - 1 = 0\}$. A probability mass function (pmf) from
$\mathcal\Pi_\tau $ at which Tsallis' (or R\'enyi's) entropy attains its
maximum will be called $\tau$-projection.

Since an argument which is presented at Section 5.4 is valid both for
Tsallis' constraints and MNNP constraints, both they will be referred
hereafter as $X$-frequency moment constraints.

\subsection{\palom 5.1 maxTent: backward compatibility with MaxEnt}

Non-extensive Thermodynamics (NET) prescribes to use maximization of
Tsallis' entropy for the pmf selection when the feasible set is defined by
$X$-frequency constraints. As it was already mentioned, the distributions
selected  by maximization of Tsallis' entropy is the same as that by
R\'enyi's entropy maximization. Though it is not our concern here, for
completeness it should be noted  that R\'enyi's entropy is extensive
(additive) whilst Tsallis' one is not, and that the 'world according to
R\'enyi' has different properties than the 'world according to Tsallis'
(see  \cite{Jizba}).

Maximization of R\'enyi-Tsallis' entropy under $X$-frequency constraints
satisfies the elementary requirement of backward compatibility with MaxEnt:
when $X$-frequency constraints reduce to the classic linear moment
constraints, the Tsallis' entropy reduces to Shannon's one (it happens for
$\alpha \rightarrow 1$). In relation to this, it should be noted that
maximization of Shannon's entropy is from the point of view of probabilistic
justifications just a special case (uniform $\vc q$) of Relative Entropy
Maximization. However no relative form of Tsallis' entropy was yet
considered by adherents of NET. For this reason in our considerations
general prior distribution $\vc q$ is replaced by uniform one, $\vc u$.

\subsection{\palom 5.2 MaxEnt: non-exponentiality\\ maxTent: power law}

Maximization of Shannon's entropy under MNNP form of $X$-frequency moment
constraints by Lagrange multiplier technique leads to pmf which is of
implicit and self-referential form, only: $p_i(\lambda) \propto e^{-\lambda
\alpha (x_i - b) p_i^{\alpha-1}}$. Whether it is the $I$-projection and
whether it is unique cannot be analytically assessed.

Under MNNP constraints, maximization of R\'enyi-Tsallis' entropy by means of
Lagrangean leads to the first order conditions for extremum which are
solved by a pmf of power-law form: $p_i(\lambda) \propto  (1 + \lambda x_i
(\alpha - 1))^{1/(1-\alpha)}$ (see \cite{MNPP}). It is important to note,
that the candidate pmf could be a (local/global) maximum only if $\alpha >
0$ and if $1 + \lambda x_i (\alpha-1) > 0$ for all $i = 1, 2, \dots, m$. The
latter requirement, known as Tsallis' cut-off condition, should be checked
on the case-by-case basis.

\subsection{\palom 5.3 Generalized entropies and BJIP}

Non-shannonian forms of entropies have been around for long time. Some of
them fall into category of convex statistical distances, and their
mathematical properties are well-studied (cf. \cite{LV}). Also, extensions
and modifications of axiomatic systems which lead to non-shannonian
entropies were studied (see \cite{AD}). Some of the 'new' entropies  were
found useful, some not (cf. \cite{Aczel}). As far as R\'enyi's entropy is
concerned few its 'operational characterizations'  were developed in the
Information Theory (cf. \cite{Arikan} and literature cited therein). Little
seems to be known however about  its probabilistic justification in context
of the ill-posed inverse problems. In particular, it is not known what is
the probabilistic question that maxTent answers. Neither it is known,
whether the unknown question which maxTent answers is meaningful to ask
within the context of BJIP.

\subsection{\palom 5.4 MaxEnt vs. maxTent}

maxTent method is by adherents of NET presented as a generalization of
MaxEnt. The generalization extends MaxEnt in two directions: Shannon's
entropy is generalized into the Tsallis' entropy, and the traditional
linear moment constraints are generalized into non-linear either Tsallis'
constraints  or MNNP constraints. Though there can be no objection made to
generalization of constraints, rather vague arguments (see for instance
Introduction of \cite{Tsallis2}) were advanced to explain why maximization
of Shannon's entropy should be under the $X$-frequency constraints replaced
by maximization of Tsallis' entropy to select a distribution from the
feasible  set which the constraints define.

Conditioned Weak Law of Large Numbers (or Gibbs conditioning principle),
Entropy Concentration Theorem and Maximum Probability Theorem provide
probabilistic justification of REM (and hence also of MaxEnt) method (though
adherents of maxTent might failed to note it, see \cite{Tsallis3}). As it
was discussed here, ECT and MPT can be readily used also under any
non-linear constraints, and hence the two Theorems give justification to
application of REM/MaxEnt also under Tsallis' or MNNP constraints. Thus,
when $n$ is sufficiently large (which is indeed the case in Statistical
Mechanics), anybody who chooses from the feasible set which is defined by
say MNNP constraints the $I$-projection(s) can be sure 1) that (any of) the
$I$-projection is just such a type in the feasible set which can be drawn
from $\vc q$ with the highest probability when $n$ goes to infinity (recall
MPT), and moreover that 2) any type which has not value of the relative
entropy close to the maximal value which is attainable within the feasible
set is asymptotically conditionally improbable (recall ECT).

In an interesting paper \cite{LS} which for the first time exposed maxTent
to a criticism from a probabilistic point of view, La Cour and Schieve
derived necessary conditions for agreement of $I$- and $\tau$-projections
under  MNNP constraints. Also, the authors illustrated by means of specific
example ($\alpha=1/2$, $m=3$, $\mathrm X = [1\  2\  3]$ and $a = 7/11$)
that $\tau$-projection can be different than $I$-projection. Provided that
the $I$-projection is unique, one can safely recall CWLLN to conclude that
maxTent-selected $\tau$-projection on $\mathcal{\Pi}_\tau$ is asymptotically
conditionally improbable. However, the issue of uniqueness or
non-uniqueness of $I$-projection on $\mathcal\Pi_\tau$ is to the best of our knowledge not
settled yet.

A different argument is used here to show that maxTent can select
asymptotically conditionally improbable distribution under $X$-frequency
constraints. The argument  is based on observation that by a choice of
support points of the random variable $X$ the feasible set of distributions
${\mathcal\Pi}_\tau$ can be made convex (the same can be done with
${\mathcal\Pi}_T$). Convexity of ${\mathcal\Pi}_\tau$ guarantees uniqueness
of $I$-projection. Provided that $\alpha > 0$ (which implies concavity of
Tsallis' entropy) the $\tau$-projection on the convex ${\mathcal\Pi}_\tau$
is as well unique. Both $I$-projection and $\tau$-projection can be then
found out by straightforward analytic maximization. Since the two are
(except of trivial cases) different, CWLLN implies that the one chosen by
maxTent has asymptotically zero conditional probability.

The next Example illustrates the argument.

{\bf Example 3}:\ \  Let $\mathcal{\Pi}_\tau = \{\vc p: \sum_{i=1}^3 p_i^2
(x_i - b) = 0,\sum_{i=1}^3 p_i - 1 = 0\}$. Let  $\mathcal{X} = [-2\ \ 0\  \
1]$ and let $b = 0$. Then $\mathcal{\Pi}_\tau = \{\vc p: p_3^2 = 2 p_1^2,
\sum p_i - 1 =0\}$ which effectively reduces to $\mathcal{\Pi}_\tau = \{\vc
p: p_2 = 1 - p_1(1 + \sqrt{2}), p_3 = \sqrt{2} p_1\}$. Prior generator $\vc
q$ is assumed to be uniform $\vc u$.

The feasible set $\mathcal{\Pi}_\tau$ is convex. Thus $I$-projection
$\hat{\vc p}$ of $\vc u$ on $\mathcal{\Pi}_\tau$ is unique, and can be
found by direct analytic maximization to be $\hat{\vc p} = [0.2748 \ \
0.3366  \ \ 0.3886]$. Straightforward maximization of R\'enyi-Tsallis'
entropy lead to unique $\tau$-projection $\hat{\vc p}_T = [0.2735 \ \
0.3398 \ \ 0.3867]$, which is different than $\hat{\vc p}$. \qed

The finding that $\tau$-projection can be asymptotically conditionally
improbable prompts Jaynes question: What are adherents of maxTent
accomplishing when they maximize R\'enyi-Tsallis' entropy?

\section{\paloa 6 Concluding comments}

Frequency moment constraints, which are  the simplest of non-linear
constraints, were employed in this work to define feasible set of types
${\mathcal\Pi}_f$ for Boltzmann-Jaynes Inverse Problem. Non-linearity of
the frequency constraints implies non-convexity of the feasible set, and
together with their symmetry also non-uniqueness of $I$-projection.
Moreover, because of the non-linearity, $I$-projections of $\mathbf q$ on
the feasible set ${\mathcal\Pi}_f$ do not take the canonical exponential
form\footnote{It obviously does not mean that they cannot be ex post
brought into the canonical exponential form. Any vector of non-negative
numbers which add up to one is MaxEnt canonical distribution, recall
\cite{Smith}, Thm. 4.1.}.

The non-linearity, non-convexity, non-uniqueness and non-exponentiality
revealed limitations of several justifications of the REM/MaxEnt method.
However, REM is not left completely unjustified in this
non-traditional setup, since two justifications of REM are provided by
Entropy Concentration Theorem and Maximum Probability Theorem. Thus though
REM under frequency constraints loses two of its charming properties:
uniqueness and exponentiality of $I$-projection, its application within the
corresponding BJIP remains justified by the two Theorems. One of the primary
aims of this work was to give a general (multiple $I$-projection)
formulation of Maximum Probability Theorem and provide its illustration. At
the same time the work was intended to  serve as an invitation to the
challenging world of non-linear constraints which shake several
traditional views of REM/MaxEnt\footnote{In particular, they call for
reconsideration of CWLLN. The law states that types conditionally
concentrate on the $I$-projection, provided that the last is unique. What
if $\mathcal\Pi$ admits several $I$-projections? Do
types concentrate on each of the $I$-projections? If yes, what is the proportion?
Answers to these questions were given elsewhere (see \cite{ggAEI}). There a
Theorem which extends CWLLN to the case of multiple $I$-projections was
stated, proven and illustrated. In order to leave the reader chance to
appreciate extent of the challenges which non-linear constraints pose to
justifications of REM/MaxEnt the present paper was intentionally written as
if the answers to these questions were not known.}.

Maximum R\'enyi/Tsallis' entropy method (maxTent) was considered here mainly
because of the non-linearity of the constraints which are used in
Non-extensive Thermodynamics (NET). As it was shown (see Sect. 5), under
the constraints maxTent can select a distribution which is according to
CWLLN asymptotically conditionally improbable. This finding prompts Jaynes
question: What are adherents of maxTent accomplishing when they maximize
R\'enyi-Tsallis' entropy? 
When it will be answered, maxTent could enter the tiny class of entropies
for which the answer is known and which can thus be consciously applied for
distribution selection.

\section{\paloa Acknowledgements}

Hospitality of Banach Centre (BC) of the Institute of Mathematics of Polish
Academy of Sciences, where a part of this study was performed as a part of
the European Community Center of Excellence programme (package 'Information
Theory and its Applications to Physics, Finance and Biology') is gratefully
acknowledged. The work was also supported by the grant VEGA 1/0264/03 from
the Scientific Grant Agency of the Slovak Republic.

It is a pleasure to thank Brian R. La Cour for very valuable discussions and
comments on \cite{ggwawa}. The thanks extend also to Ale\v s Gottvald,
George Judge, Jonathan D. H. Smith and Viktor Witkowsk\'y.

\appendix

\section{\paloa Appendix}

Observe, that any of the three $I$-projections at the Example 2 (Section
4.2)  has two of probabilities equal. This can be elucidated by the
following elementary considerations: suppose that  the feasible set is
constrained further by additional requirement $p_1 = p_2 = p_3$. This
additional requirement makes $\vc p_{0} = [1/3\ \,1/3\ $ $1/3]$ the only
pmf in the set. Clearly, the pmf is indeed in the set only if $a \equiv a_0
= 1/3$, ie. the 'centre of mass' of $\sum_{i=1}^3 p_i^2$. If $a \neq a_0$
then $\vc p_{0}$ is not in $\mathcal{\Pi}_f$, hence the most entropic pmf
should be sea\-rch\-ed am\-ong tho\-se pmf's which have two of
probabilities equal; say $p_1 = p_2$.

The additional requirement turns the under-de\-ter\-mi\-ned
con\-di\-ti\-ons in\-to a qua\-dra\-tic e\-qua\-ti\-on which is solved by
either $p_1 = 0.2131$ or $p_1 = 0.4535$. Hence the restricted feasible set
comprises two groups of pmf's $[0.2131\ 0.2131\ 0.5737]$ and $[0.4535\
0.4535$ $0.0930]$. The first pmf has Shannon's entropy $H_U = 0.9777$, the
second $H_L = 0.9381$. It does not surprise that pmf's from the original set
$\mathcal{\Pi}_f$ (ie. those which can have all three probabilities
different) have Shannon's entropy within the bounds which are set up by
$H_L$ and $H_U$.

This is obviously, not a property specific to the studied example with the
particular choice of $\alpha = 2$ and $m=3$. In general, the finding
permits to state the following

{\bf Proposition}\ \ \ {\it Let $\vc q$ be uniform, $\mathcal{\Pi}_f
\triangleq \{\vc p: \sum p_i^\alpha - a = 0, \sum p_i - 1 = 0\}$, where
$\vc p \in \mathbf{R}^m$ and  $\alpha \in \mathbf{Z}$. Let $m > \alpha$.
Then $\hat{\vc p} \in \mathcal{\Pi}_f$ such that
 $H(\vc p) \le H(\hat{\vc p})$ for any $\vc p \in \mathcal{\Pi}_f$, is such that
$\hat{p}_1 = \hat{p}_2 = \dots = \hat{p}_{m-1}$, where $\hat{p}_1$ is one
of solutions of the following algebraic equation: }
\begin{equation}
 (m-1)\hat{p}_1^\alpha + (1-(m-1)\hat{p}_1)^\alpha - a = 0
\end{equation}

Note: Clearly, among the pmf's which solve equation (3), $\hat{\vc p}$ is
the one with  the highest value of Shannon's entropy $H$. Any permutation of
$\hat{\vc p}$ is also $I$-projection of $\vc u$ on $\mathcal\Pi_f$.

\section{\paloa Bibliographic note}

Literature on Tsallis' maximum entropy method is vast (cf. \cite{bibl}).
arXiv contains a series of preprints which document evolution of the
method. Also, see March 2002 issue of Chaos, Solitons and Fractals.
Interesting  introductory remarks on NET can be found at \cite{Cohen}.
Critical voices are rare: besides the fundamental \cite{LS} see for
instance also \cite{Gottvald}, \cite{Velasquez}. This work draws on and
corrects \cite{ggwawa}.

\bigskip\medskip

 \leftline{July 2002, May-July 2003}

\medskip

\renewcommand\refname

 \bigskip\bigskip

 \rightline{\palomar in memory of El Mar}

\end{document}